\documentclass[amsmath,amssymb,aps,floats,prx,twocolumn,showpacs,superscriptaddress,longbibliography]{revtex4-1}
\usepackage{float}
\usepackage[final]{graphicx}
\usepackage{amssymb,amsmath}
\usepackage{color}
\usepackage{xcolor}
\usepackage{longtable}
\usepackage{graphicx}
\usepackage{epstopdf}
\usepackage{dcolumn}
\usepackage{bm}
\usepackage{tabularx}
\usepackage[subnum]{cases}
\usepackage[dvipsnames]{xcolor}
\begin{document}

\preprint{APS/123-QED}

\title{Terahertz light driven coherent excitation of a zone-folded Raman-active phonon mode in the Spin-Ladder System $\alpha'$-NaV$_2$O$_5$}

\author{Flavio Giorgianni}
\email{flavio.giorgianni@unibe.ch}
\affiliation{Institute of Applied Physics, University of Bern, CH-3012 Bern, Switzerland}
\author{Martina Romani}%
\affiliation{INFN-LNF, via E. Fermi 40, Frascati, 00044, Italy}%
\author{Pascal Puphal}%
\affiliation{Max Planck Institute for Solid State Research, Heisenbergstrasse 1, 70569 Stuttgart, Germany}%
\author{Masahiko Isobe}%
\affiliation{Max Planck Institute for Solid State Research, Heisenbergstrasse 1, 70569 Stuttgart, Germany}%
\author{Leonie Spitz}%
\affiliation{Paul Scherrer Institute, CH-5232 Villigen-PSI, Switzerland}%
\author{Mariangela Cestelli Guidi}%
\affiliation{INFN-LNF, via E. Fermi 40, Frascati, 00044, Italy}%
\author{Carlo Vicario}%
\affiliation{Paul Scherrer Institute, CH-5232 Villigen-PSI, Switzerland}%
\author{Mattia Udina}%
\affiliation{ISC-CNR and Department of Physics, “Sapienza” University of Rome, P.le Aldo Moro 5, 00185, Rome, Italy}

\date{\today}

\begin{abstract}
We investigate the out-of-equilibrium lattice dynamics in the spin-ladder system $\alpha'$-NaV$_2$O$_5$ using intense terahertz (THz) pump and near-infrared (NIR) probe spectroscopy. When quasi-single-cycle THz pulses interact with $\alpha'$-NaV$_2$O$_5$ in its low-temperature, dimerized charge-ordered phase, they induce coherent oscillations in the time domain at the zone-folded Raman-active phonon frequency of 1.85 THz. By combining pump-probe measurements with lattice dynamics modeling based on equation-of-motion approach, we propose that these oscillations arise from a nonlinear coupling between Raman-active and infrared (IR)-active phonon modes, with the latter being resonantly excited by the THz pulses. In contrast, excitation with NIR femtosecond laser pulses does not produce measurable vibrational dynamics, highlighting the unique potential of THz-driven, nonlinear light-matter interactions for the coherent and selective control of structural dynamics in quantum materials.
\end{abstract}

\maketitle

\section{Introduction}
Studies of ultrafast phenomena in complex materials have been revolutionised by recent advances in coherent light sources that have enabled the generation of high-intensity, ultrashort light pulses at specified THz and mid-infrared (MIR) frequencies~\cite{salen2019,zhang2021,zong2023}. These sources have proven to be highly effective in the dynamic manipulation of electronic, magnetic, and structural properties on femtosecond and picosecond timescales~\cite{RevModPhys.93.041002,basov2017,disa2023}. Driving materials far from equilibrium with tailored THz and MIR fields, has revealed – both experimentally and theoretically – a rich landscape of metastable and nonlinear phenomena, transient quantum phases~\cite{disa2021,budden2021,buzzi2020,li2019terahertz,wang2013,basini2024}, and novel pathways for the selective excitation of collective modes such as lattice vibrations~\cite{juraschek2018,juraschek2021,Maehrlein2017,johnson2019,giorgianni2022}, spin waves~\cite{afanasiev2021,giorgianni2023,mashkovich2021} and collective modes in superconductors~\cite{vaswani2021,schwarz2020,chu2020}.

THz and MIR pulses, are uniquely suited for driving low-energy collective excitations that are otherwise inaccessible via direct light-matter coupling. Through nonlinear coupling mechanisms, such as phonon-phonon, electron-phonon, and spin-phonon interactions, these pulses can be used to selectively manipulate individual degrees of freedom without introducing significant heating in the material~\cite{RevModPhys.93.041002,basov2017,disa2023}. This selective excitation permits precise control of strongly correlated materials, where the interplay among charge, spin, and lattice dynamics gives rise to macroscopic quantum phenomena, including multiferroicity, unconventional superconductivity, and charge or spin ordering~\cite{keimer2017}.

Here, we investigate the ultrafast dynamics of Raman-active zone-folded phonon modes in the prototypical spin-ladder system $\alpha'$-NaV$_2$O$_5$, which is driven by nonlinear THz light-matter interaction. An intense THz pulse induces coherent oscillations at the frequency of a Raman-active zone-folded phonon mode ($\Omega_R = 1.85$ THz). Temperature-dependent measurements show that these oscillations vanish during the structural phase transition, confirming their origin as zone-folded lattice vibrations associated with the low-temperature, charge-ordered dimerized phase. The quadratic dependence of the oscillation amplitude on the THz field strength also signals a nonlinear excitation mechanism. As the coherent dynamics are only evident when the spectral components of the THz pulse overlap with IR-active phonon frequencies, it can be concluded that the excitation mechanism is primarily governed by nonlinear lattice coupling between IR-active phonons and the Raman-active phonon. More specifically, interpretation of the results with phonon dynamics modeling based on coupled equations of motion suggests that the nonlinear excitation of the Raman-active zone-folded phonon mode is due to driving IR-active phonons with intense THz light as opposed to purely photonic nonlinear pathways such as frequency mixing of the THz light components. 
Importantly, femtosecond NIR laser excitation fails to induce any measurable coherent vibrational dynamics, highlighting the unique capability of THz-driven nonlinear light-matter interactions to selectively excite Raman-active phonons inaccessible through direct optical coupling.

\begin{figure*}
\includegraphics[width=1\linewidth]{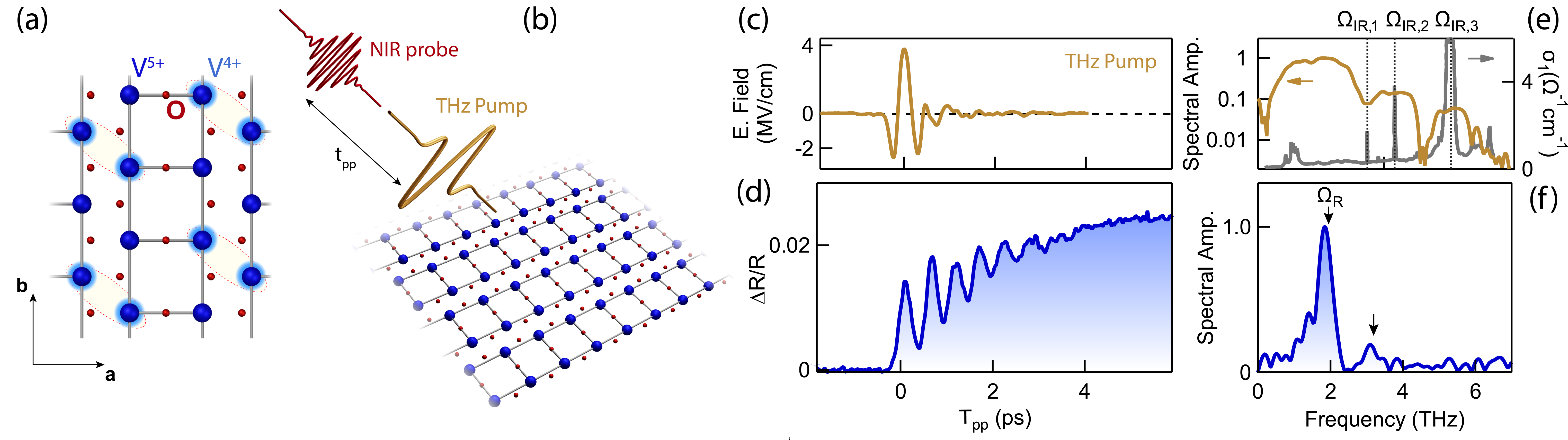}
\caption{(a) Schematic representation of the spin-ladder system formed by the Vanadium (V) and oxygen atoms (O). Below $T_c=34 K$, the system is in a charge-ordered phase, with in-equivalent $V^{4+}$ and $V^{5+}$. (b) Experimental setup. An intense THz pulse (yellow) with the electric field polarized linearly along the spin-ladder (b-axis) of $\alpha'$-NaV$_2$O$_5$ excites the crystal and induces a polarization change on a femtosecond NIR probe pulse (red). (c-d) THz electric field, measured by electro-optic sampling (EOS), and induced polarization changes of the probe as a function of the time delay $t_{pp}$. (e) Fourier amplitude spectrum of the THz pump waveform (yellow solid line) and real part of the optical conductivity $\sigma_1$ along the $b$-axis of $\alpha'$-NaV$_2$O$_5$ at 4.4 K, taken from~\cite{room04}. (f) Fourier transform of the temporal derivative of $\Delta R/R$ shown in panel (d). Black arrows indicate the frequencies of the Raman-active modes from Ref.~\cite{lemmens1998}.}
\label{tnpfig1}
\end{figure*}

\section{The Experiment}
The spin-ladder system $\alpha'$-NaV$_2$O$_5$ experiences an intriguing phase transition characterized by charge ordering, spin pairing, and lattice distortion at the critical temperature $T_c\sim 34$ K~\cite{ohama1999}. Above $T_c$, the vanadium (V) ions form ladder structures in the $a$-$b$ plane and have an average +4.5 charge. Below $T_c$, a charge ordering transition occurs, leading to an alternating arrangement of V$^{4+}$ and V$^{5+}$ ions along the $b$ axis, as depicted in Fig.~\ref{tnpfig1}(a). This charge order phase leads neighboring V$^{4+}$ ions to form spin dimers in a spin-singlet state, causing the opening of a spin gap in the magnetic spectrum~\cite{ohama1999,smolinski1998}.

Simultaneously with the formation of the charge-ordered phase, a lattice superstructure emerges due to lattice distortion accompanied by a spin dimerization~\cite{popova2002,popova2002_2}. This new lattice configuration, in turn, leads to the appearance of spectrally narrow IR- and Raman-active phonon modes in the THz range due to Brillouin-zone folding~\cite{popova2002,popova2002_2}. Raman spectroscopy has identified a low-frequency zone-folded phonon mode at $\Omega_R$ = 66 cm$^{-1}$ (1.85 THz), emerging as the primary contribution to the Raman scattering cross-section in the $a$-$b$ plane~\cite{lemmens1998,konstantinovic2001,suemoto2005,nakajima2005}. In contrast, two additional zone-folded phonon modes at 107 cm$^{-1}$ (3.20 THz) and 134 cm$^{-1}$ (4.02 THz) exhibit a comparatively lower Raman scattering cross-section. Initially assigned to a magnetic excitation due to the comparable energy of the spin-gap~\cite{lemmens1998,nakajima2005}, further investigations have revealed the vibrational nature of the mode at $\Omega_R$ through temperature and doping-dependent analyses employing Raman, IR, and ESR spectroscopies~\cite{popova2002,popova2002_2,room04,luther1998}. However, it is important to acknowledge that the potential involvement of spin interactions in the vibrational dynamics of this mode cannot be categorically excluded \textit{a priori}. Additionally, Raman studies, coupled with theoretical considerations, suggest that charge order fluctuations, modulating the spin super-exchange interactions, can result in a strong spin-phonon coupling below T$_c$~\cite{sherman1999}. This spin-phonon coupling has been suggested to significantly enhance vibrational anharmonicity~\cite{sherman1999}, making $\alpha'$-NaV$_2$O$_5$ a
compelling system for investigating nonlinear THz-driven out-of-equilibrium dynamics and exploring potential nonlinear phononic mechanisms.

Infrared spectroscopy in the THz range, polarized along the $b$ axis, has revealed the emergence of IR-active, zone-folded phonon modes below T$_c$~\cite{popova2002,room04}. These modes appear at $\Omega_{IR,1}$ = 101 cm$^{-1}$ (3.03 THz), $\Omega_{IR,2}$ = 127 cm$^{-1}$ (3.8 THz), and are associated with the formation of the lattice superstructure~\cite{popova2002,room04}. Additionally, a phonon mode is observed at $\Omega_{IR,3}$ = 180 cm$^{-1}$ (5.4 THz)~\cite{popova2002,room04}. Other phonons have a weaker IR absorption cross-section, including the low-frequency phonon cluster at around 35 cm$^{-1}$ ($\sim$ 1 THz)~\cite{room04}. All the zone-folded phonon modes rapidly disappeared above $T_c$.\\ 

For our experiment, high-quality single crystals of $\alpha'$-NaV$_2$O$_5$ were grown by self-flux method~\cite{isobe1997}. The high crystal quality was confirmed by Laue diffraction. The samples were cut and polished on the $a$-$b$ plane with dimensions of the order of 2 x 1.5 mm$^2$. To explore the potential for driving coherent structural dynamics in $\alpha'$-NaV$_2$O$_5$, we performed ultrafast THz pump and NIR probe spectroscopy, illustrated in Fig.~\ref{tnpfig1}(b). Intense quasi-single-cycle THz pulses were focused onto the sample mounted on a cold finger of a cryostat. The THz-induced response was then tracked by measuring the reflected intensity in a quasi-collinear configuration using ultrashort ($\sim$ 55 fs) NIR laser probe pulses, at a photon energy of 1.55 eV, as a function of the pump-probe time delay $t_{pp}$ [see Supplemental Material~\cite{supp}, which includes Refs.~\cite{Giorgianni2019,vicario2020,Blanchard2009,khalsa2021,huber2015,Bistoni2019,konstantinovic1999,XrayNVO,damascelli2000,xray}.).

The THz-induced response was monitored by measuring the optical reflectivity modulation $\Delta R/R=(R_a-R_b)/R$. $R_a$ and $R_b$ denote the reflected probe intensities under THz pumping along the $a$ and $b$ axes, respectively, and $R=R_a+R_b$ represents the reflected intensity without the THz pump pulse. The $\Delta R/R$ response can effectively capture the coherent lattice dynamics of the Raman-active phonon mode at frequency $\omega_R$, since its phonon displacement alters the polarization state of the NIR probe, which is supported by its activity in the cross-polarized ($ab$) Raman scattering channel~\cite{lemmens1998, popova2002}.

The probe pulses were circularly polarized to allow a more direct comparison with previous optical pump-probe studies~\cite{kamioka2002,suemoto2005}. Additionally, we verified that the THz-induced dynamics remain unchanged by probing it with linear polarization at 45$^\circ$ with respect to the $b$ axis, see Sec. S1 of Supplemental Material~\cite{supp}.

The temporal waveform of the THz pump electric field $E(t)$, see Fig.~\ref{tnpfig1}(c), was measured at the sample position through electro-optic sampling utilizing a 200 $\mu$m-thick GaP crystal (see Sec. S1 of the Supplemental Material~\cite{supp}). In Fig.~\ref{tnpfig1}(d), the temporal evolution of the THz pump-induced $\Delta R/R(t_{pp})$ is depicted. The time origin ($t_{pp}=$0) is conventionally defined at the delay when the optical probe overlaps the peak of $E(t)$.
The observed THz-induced dynamics exhibit coherent oscillations superimposed on a gradually increasing background. It is important to emphasize that the envelope function of the oscillatory response extends well beyond the time delays when the amplitude of the THz field becomes negligible. These long-lasting oscillatory dynamics imply that the oscillations result from a coherent excitation of a vibrational mode.
\begin{figure}
\includegraphics[width=1\linewidth]{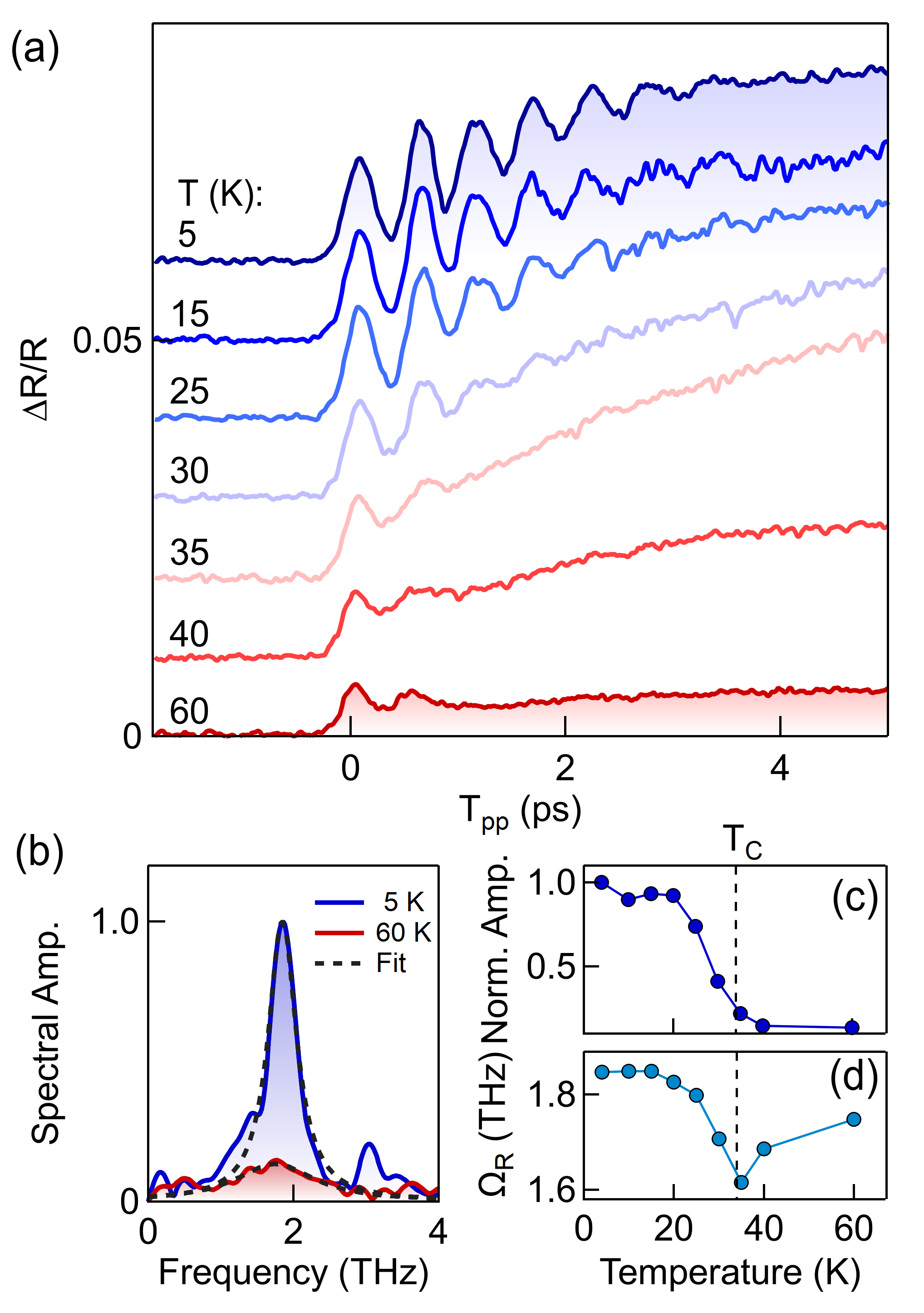}
\caption{\label{tnpfig2} (a) Temperature evolution of the THz pump-induced $\Delta R/R$ in $\alpha'$-NaV$_2$O$_5$. No prominent signature of coherent oscillations is observed above $Tc=34 K$. Vertical offsets have been added to the curves for better visualization. (b) Fourier transform of $d(\Delta R/R)/dt_{pp}$ for T= 5 and 60 K. Black dotted curves are Lortentzian fit, to evaluate the amplitude and the center mode frequency $\Omega_R$. (c-d) Normalized amplitude and center frequency $\Omega_R$ as a function of temperature determined by Lorentzian fit.}
\end{figure}
For a more comprehensive analysis of the THz-induced dynamics and their underlying frequency components, we employ the Fourier transform of both the THz pump waveform and the temporal derivative of the dynamics, represented as $d(\Delta R/R)/dt_{pp}$. The use of the temporal derivative is commonly adopted in ultrafast spectroscopy to mitigate the quasi-DC effect arising from slowly varying background signals in the pump-probe time traces~\cite{kamioka2002}.
Fig.~\ref{tnpfig1}(e) shows the THz pump spectrum compared with the real part of the optical conductivity along the $b$ axis, which shows three main phonon excitations at $\Omega_{IR,1}$, $\Omega_{IR,2}$, and $\Omega_{IR,3}$.
As illustrated in Fig.~\ref{tnpfig1}(f), the Fourier amplitude unveils a distinct peak at 1.85 THz, corresponding to the frequency of the zone-folded Raman-active phonon mode at $\Omega_R$. Furthermore, the presence of the weaker Raman-active phonon around 3.02 THz (107 cm$^{-1}$) is also detected~\cite{popova2002}.

To confirm that the observed oscillatory dynamics actually arise from the zone-folded phonons, we investigated the evolution of $\Delta R/R(t_{pp})$ dynamics across the temperature-driven phase transition. In Fig.~\ref{tnpfig2}(a), the THz-induced $\Delta R/R(t_{pp})$ dynamics are presented for temperatures ranging from 5 K to 60 K. Notably, as the temperature approaches $T_c$, coherent oscillations rapidly vanish while the contribution of the slowly varying background becomes dominant. In Fig.~\ref{tnpfig2}(b), the Fourier transform amplitude of $d(\Delta R)/dt_{pp}$ for two different temperatures ($T= 5 K<T_c$ and $T= 60 K>T_c$) is illustrated, indicating the disappearance of the peak at $\Omega_R$ for $T>T_c$. Moreover, the Fourier amplitude and the peak frequency as a function of temperature, obtained by fitting procedure using a Lorentzian function, are shown in Fig.~\ref{tnpfig2}(c)-(d).
\begin{figure}
\includegraphics[width=1 \linewidth]{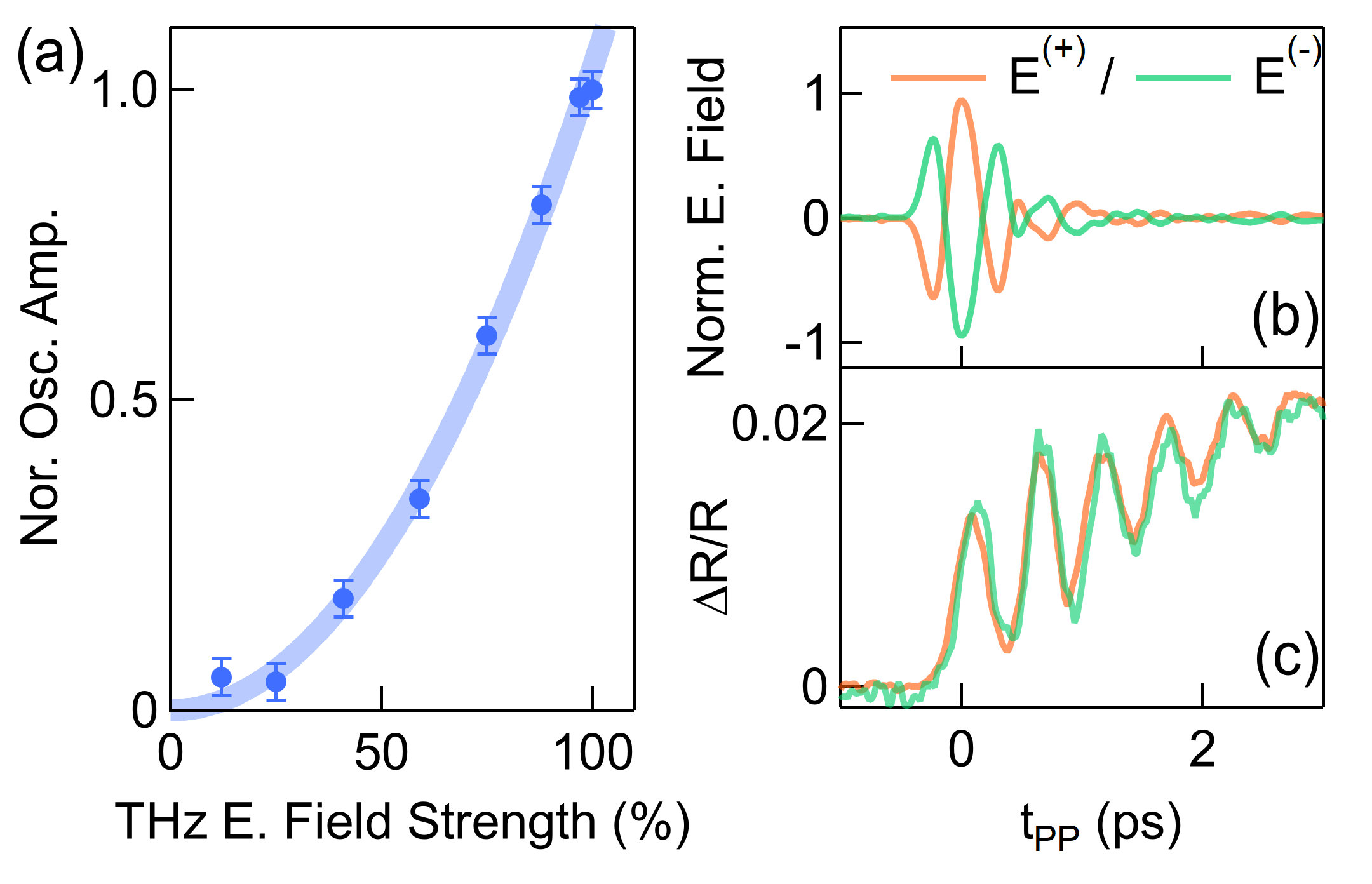}
\caption{(a), Normalized oscillations amplitude (Norm. Osc. Ampl.) vs normalized THz field strength. The amplitude scales quadratically with the pump electric field (the light blue solid line represents a quadratic fit). Error bars indicate the standard deviations of the mean determined from pump-probe scan measurements, (b) THz pump electric field waveform $E^{(+)}$ (dashed line) and polarity inverted one $E^{(-)}$ and corresponding THz induced $\Delta R/R(t_{pp})$ dynamics.}
\label{tnpfig3}
\end{figure}

To gain a comprehensive understanding of the driving mechanism of the observed phonon excitation, we conducted measurements to examine the relationship between the oscillation amplitude of the driven oscillatory dynamics and the electric field amplitude of the incident THz field, see Fig.~\ref{tnpfig3}(a). The measured dynamics show a clear quadratic trend, providing compelling evidence for the nonlinear coupling of the mode to the driving electric field. Additionally, we exclude the possibility of a linear (dipolar) coupling of light to the observed coherent mode, since the mode is purely Raman active. As supporting evidence, Fig.~\ref{tnpfig3}(b) demonstrates that the dynamics do not change upon polarity inversion of the THz field, providing a strong indication of no direct dipolar excitation.
\begin{figure*}
\includegraphics[width=1.05\linewidth]{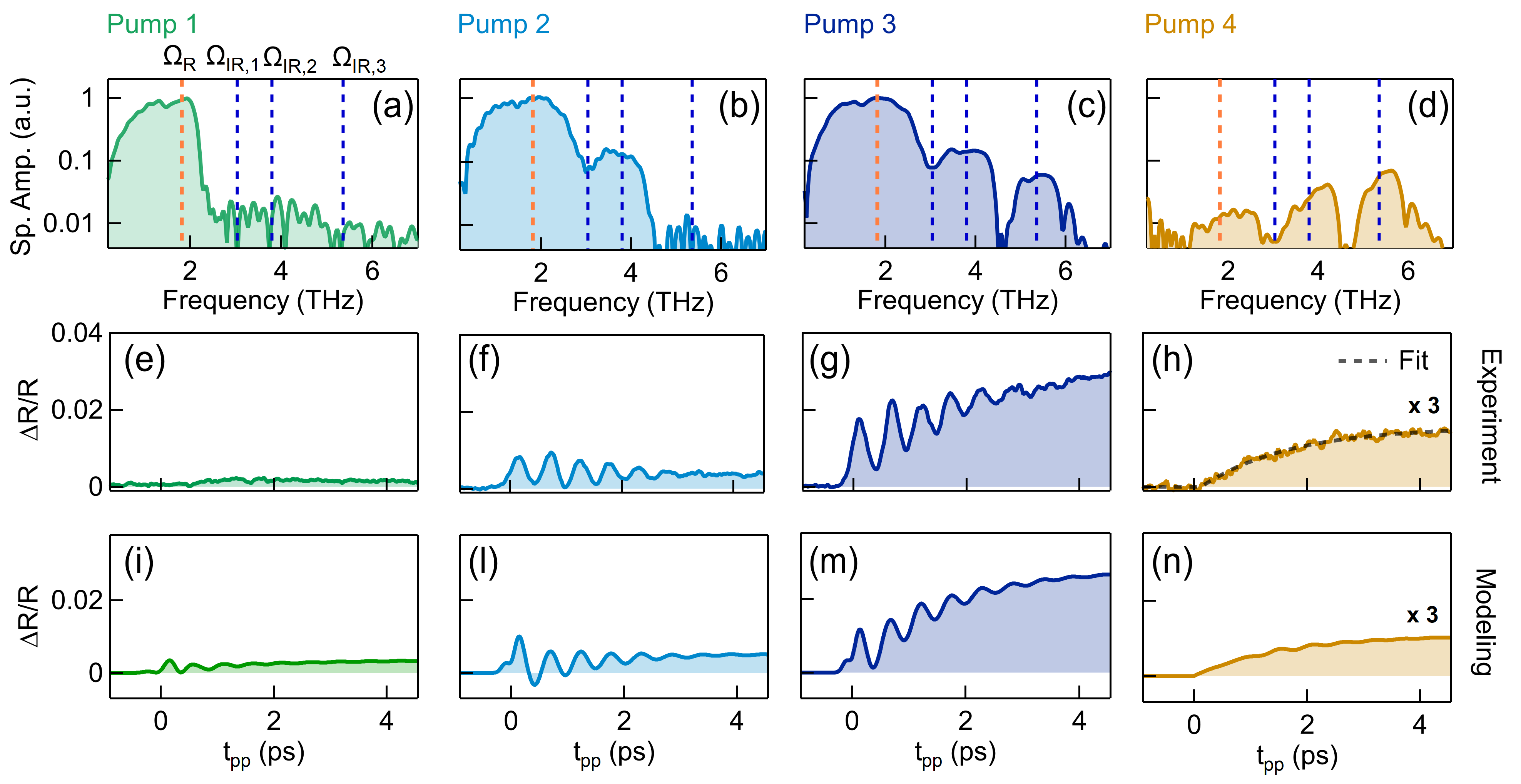}
\caption{\label{tnpfig4} (a)-(d), THz pump excitation spectra. The dotted lines indicate the resonance frequency of the Raman phonon, $\Omega_R$, and IR-active phonons, $\Omega_{IR,1}$, 
$\Omega_{IR,2}$, and $\Omega_{IR,3}$. (e)-(f), Corresponding THz-induced $\Delta R/R$ dynamics. $\alpha'$-NaV$_2$O$_5$ is kept at 5 K. The black dashed line in panel (h) is the exponential fit, see main text. (i)-(n), Numerical calculation using Eq. (4) based on nonlinear phononic model.}
\end{figure*}

Next, we discuss the nonlinear excitation mechanism of the vibrational modes by THz driving that we observed. We conducted a series of THz pump - NIR probe experiments, employing THz spectral filters to tailor the spectrum of the THz driving field, see Fig.~\ref{tnpfig4}(a)-(d). Dotted lines indicate the frequencies of the zone-folded Raman and IR active phonons.

When the THz spectrum remains below 2.5 THz, our measurements reveal no significant oscillatory dynamics, see Fig.~\ref{tnpfig4}(e), despite the overlap between the THz pump spectral frequencies with the Raman-active phonon mode frequency at $\Omega_R$. This observation strongly suggests that either linear (dipolar) excitation or nonlinear excitation pathways purely involving photons are very unlikely. Extending the spectral contents up to 4.5 THz, see Fig.~\ref{tnpfig4}(b), the THz pump is able to produce the oscillatory vibrational dynamics as the pump frequency contents match those of the IR-active phonons at $\Omega_{IR,1}$=3.03 THz and at $\Omega_{IR,2}$ = 3.8 THz. Superimposed on the oscillatory dynamics, we also observe a concomitant rise of the slowly varying background. Further extending the THz pump spectral components up to 6 THz, see Fig.~\ref{tnpfig4}(c) reaching the frequency of the phonon at $\omega_{IR,3}$ = 5.4 THz, resulted in a substantial increase in the background.
Finally, when the pump spectral contents are centered mainly around $\omega_{IR,3}$, only the slowly varying component in the dynamics is observable, see Fig.~\ref{tnpfig4}(d). The temporal dynamics are accurately reproduced by a single exponential function $(1-e^{-t_{pp}/\tau})$, a characteristic often associated with thermal processes~\cite{Giorgianni2019,suemoto2005}. The fitting of this function yields a time constant of $\tau$ = 1.58 ps.
Remarkably, as one can see in Fig.~\ref{tnpfig4}, the absence of oscillatory dynamics in cases where the pump spectral contents lie below 2.5 THz (Pump 1) and above 4 THz (Pump 4) strongly suggests that coherent oscillations of the Raman mode are driven through a nonlinear phononic process involving the excitation of the phonon modes at $\Omega_{IR,1}$, $\Omega_{IR,2}$, and $\Omega_{IR,3}$.\\

\section{Nonlinear Phononics as a Possible Excitation Mechanism}
In the following, we model the lattice dynamics driven by nonlinear phononic excitation using an equation-of-motion approach. The indirect excitation of a coherent Raman mode, which arises from the resonant large-amplitude IR-active vibrations, can be described through the lattice potential to the lowest order beyond the harmonic approximation~\cite{juraschek2018,martin1974}:
\begin{equation}\label{eq1}
V=\frac{\Omega_R^2 Q_R^2}{2}+\sum_{i}\frac{\Omega_{IR,i}^2 Q_{IR,i}^2}{2}-\sum_{i,j}c_{ijR} Q_{IR,i} Q_{IR,j} Q_R.
\end{equation}
Here, $Q_{R}$ and $\Omega_{R}$ represent the phonon normal coordinate and the eigenfrequency of the Raman-active phonon mode, respectively, while the indices $i,j= 1,2,3$ denote those of the IR-active modes. The parameters $c_{ijR}$ are the nonlinear coupling coefficients. 

Given that the IR-active phonons are directly excited to a large amplitude, the phonon coordinate $Q_{IR,i}$ is significantly greater than $Q_{R}$, i.e., $Q_{IR,i} \gg Q_{R}$. This allows us to neglect the coupling terms $c_{iiR} Q_{IR,i} Q_R^2$.
Consequently, the relevant nonlinear anharmonic coupling to the lowest order is given by the term $Q_{IR,i} Q_{IR,j} Q_R$.

The lattice energy $U$ is given by $U = V - \sum_{i} Q_{IR,i} Z_i E$, which accounts for the interaction between the IR-active phonon modes and the electric field of the THz pump pulse $E(t)$~\cite{melnikov2018,juraschek2018}. Here, $Z_i$ represents the phonon effective charge of the $i$ phonon mode. 

Following the approach in Refs~\cite{melnikov2018,juraschek2018}, we derive the equations of motion for both the Raman and IR-active phonons from the lattice potential $U$:

\begin{numcases}{}
\ddot{Q}_{IR,i}+ 2 \gamma_{IR,i} \dot{Q}_{IR,i}+\Omega_{IR,i}^2 Q_{IR,i} + \nonumber \\
-\sum_{j} c_{ijR} Q_{IR,j}  Q_{R} = Z_{i} E \label{eq2a} \\
\ddot{Q}_{R}+ 2 \gamma_{R} \dot{Q}_{R}+\Omega_R^2 Q_{R} = \sum_{ij} c_{ijR} Q_{IR,i}Q_{IR,j}
\label{eq2b}
\end{numcases}

The Eq.~(\ref{eq2b}) describes the temporal evolution of the Raman phonon coordinates $Q_{R}$ as a harmonic oscillator driven by the term $Q_{IR,i}Q_{IR,j}$, which is quadratic in the IR-active phonon coordinates. In the presence of nonlinear coupling, this term becomes relevant when IR active phonon modes are driven to large amplitude by the intense and resonant light field. To model the dynamics of the Raman active mode $Q_R$, we numerically solve Eq.~(\ref{eq2a}) and Eq.~(\ref{eq2b}), by employing the experimentally measured THz pump fields and the phonon parameters evaluated by infrared and Raman spectroscopy: $\Omega_R$ = 1.85 THz, $\gamma_R$ = 0.15 THz, $\Omega_{IR,1}$ = 3.03 THz, $\gamma_{IR,1}$ = 0.01 THz,$\Omega_{IR,2}$ = 3.8 THz, $\gamma_{IR,2}$ = 0.01 THz, $\Omega_{IR,3}$ = 5.4 THz, $\gamma_{IR,3}$ = 0.14 THz~\cite{room04,damascelli2000}. The effective charge $Z_i$ can be estimated from the real part of the optical conductivity $\sigma_1$ at the IR frequency $\omega_{IR,i}$, as detailed in Sec. S4 the Supplemental Material~\cite{supp}. In the numerical calculation, $\sigma_1(\omega_{1})$, $\sigma_1(\omega_{2})$, $\sigma_1(\omega_{3})$ are taken from the optical conductivity in Fig.~\ref{tnpfig1}, see Refs.~\cite{room04,damascelli2000} for details.

We model the induced reflectivity changes as~\cite{huber2015}:
\begin{equation}\label{eq4}
\frac{\Delta R}{R}(t_{pp})=A\cdot Q_R(t_{pp})+B\cdot\eta(t_{pp}),
\end{equation}
where the first term is proportional to the phonon coordinate $Q_R$, as discussed in Ref.~\cite{huber2015}, representing the excited nonlinear vibrational dynamics. The second term characterizes the slowly varying background, attributed to the thermal effect due to the deposition of optical energy. To model these thermal dynamics, it is reasonable to assume a single exponential form $\eta(t_{pp})=(1-e^{-t_{pp}/\tau_{THz}})$, consistent with our experimental observations, as depicted in Fig. 4(h). For the model, we set a time constant of $\tau_{THz}$ = 1.58 ps as determined by fitting the experimental data, see Fig. ~\ref{tnpfig4}(h).

\begin{figure}
\includegraphics[width=1\linewidth]{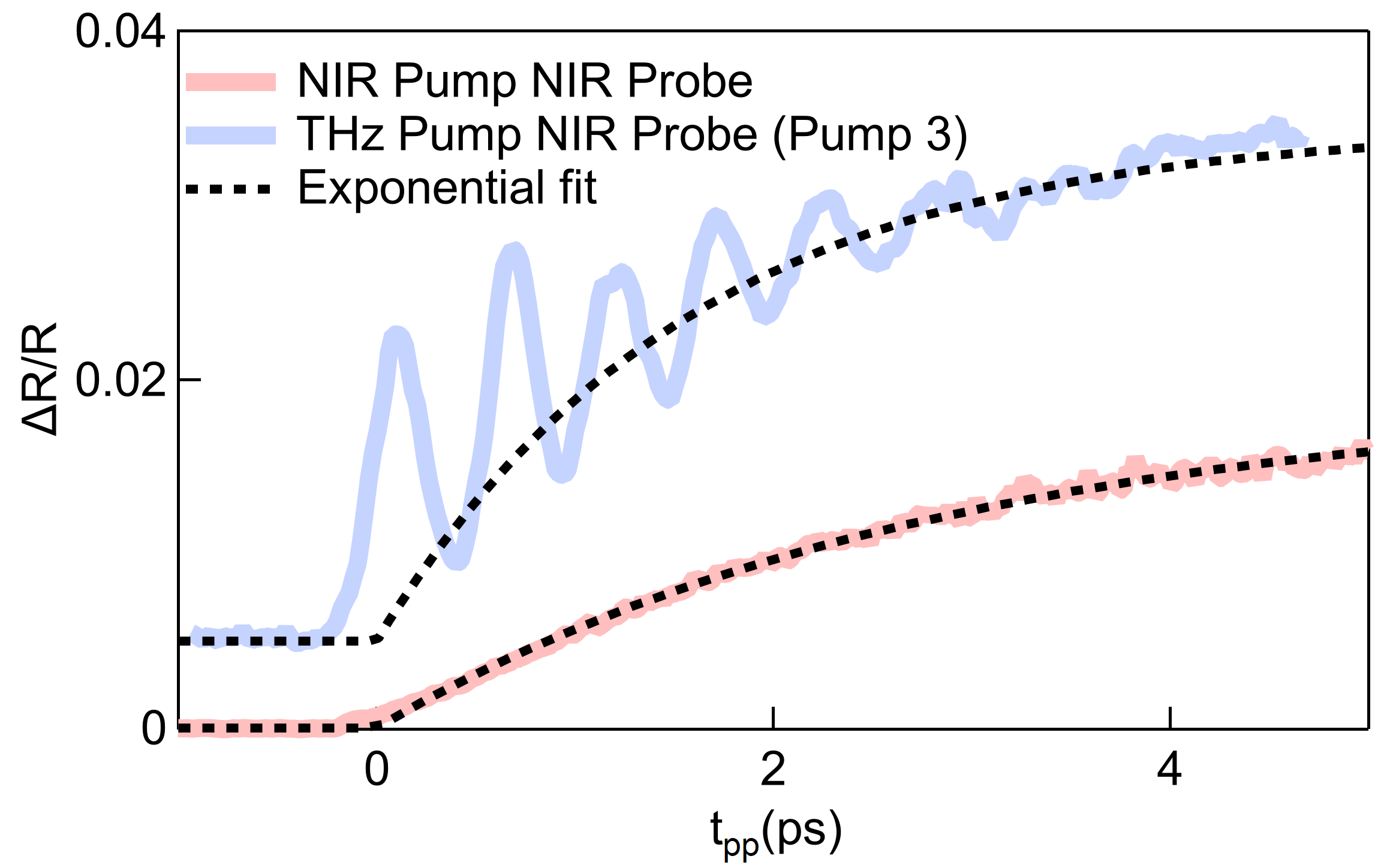}
\caption{\label{tnpfig5} Comparison between THz pump and NIR pump induced $\Delta R/R(t_{pp})$ in $\alpha'$-NaV$_2$O$_5$. The sample is kept at 5 K. NIR pump fluence is $\sim$ 3 mJcm$^{-2}$. The dotted black lines are single-exponential fit. Vertical offset has been added for better visualization.}
\end{figure}

Numerical results of Eq.~(\ref{eq4}) are reported in Fig.~\ref{tnpfig4}(i)-(n), while details about the $A$ and $B$ parameters are reported in Sec. S3 of the Supplemental Material~\cite{supp}. Our calculations show a consistent agreement with observed dynamics and elucidate its dependence on the spectral content of the THz pump. Furthermore, in Sec. S2 of the Supplemental Material~\cite{supp}, we have extended our analysis by performing a comparative study of the numerical results of nonlinear phononics with other excitation mechanisms, such as nonlinear photonics and infrared resonant Raman scattering~\cite{juraschek2018, Maehrlein2017, khalsa2021}. This analysis provides strong evidence that nonlinear phononics is the dominant driving mechanism underlying the observed dynamics. However, determining the selection rules for the nonlinear phononic excitation based on phonon symmetry compositions remains challenging due to unresolved ambiguities in the crystal symmetry of the low-temperature charge-order phase of $\alpha'$-NaV$_2$O$_5$. Competing structural models, including centrosymmetric \textit{C2/c}, noncentrosymmetric \textit{Fmm2}, and monoclinic \textit{$C_2^3$-A112} space groups, have been proposed based on infrared and Raman spectroscopy and X-ray diffraction studies~\cite{konstantinovic1999,popova2002}. If the structure is centrosymmetric (\textit{C2/c}) as discussed in Ref.~\cite{konstantinovic1999,popova2002}, the nonlinear coupling follows the composition rule B$_u$ × B$_u$=A$_g$, allowing IR phonons (B$_u$ symmetry) to excited the zone-folded Raman-active phonon (A$_g$ symmetry). Same conclusion is found for the monoclinic \textit{$C_2^3$-A112} case. However, if the crystal symmetry is non-centrosymmetric (\textit{Fmm2}), as discussed in Refs.~\cite{popova2002}, conventional composition rules cannot explain the nonlinear phononics excitation. Detailed discussion for both cases is given in Sec. S5 of the Supplemental Material~\cite{supp}.

Further complexity arises from charge order, which can influence the local crystal symmetry. Infrared and X-ray studies have revealed coexisting ladder-plane configurations characterized by distinct diagonal charge-order patterns~\cite{room04,xray}, leading to local structural variations that deviate from the average symmetry. These local symmetry changes have been proposed as the origin of the anomalous symmetry rules observed for the zone-folded IR phonons (including those at $\Omega_{IR,1}$ and $\Omega_{IR,2}$) in polarization-resolved infrared spectroscopy~\cite{room04,damascelli2000}. Consequently, these variations may alter the effective selection rules governing nonlinear phononic coupling. Further discussion on these structural and symmetry-related effects can be found in Sec. S5 of the Supplemental Material~\cite{supp}.
Nonetheless, despite these ambiguities, the remarkable agreement between the nonlinear phononic model and the experimental data - especially in contrast to models based on alternative excitation mechanisms - strongly supports nonlinear phononics as the most likely excitation pathway for the observed phenomena (see Sec. S3 of the Supplemental Material for a full analysis~\cite{supp}).

Next, we phenomenologically discuss the origin of the slowly varying background in the observed dynamics. The slow modulation of the $\Delta R/R$ change at the probe photon energy (1.55 eV) under THz excitation is indicative of a temperature increase, consistent with previous steady-state polarization-resolved infrared spectroscopy investigations~\cite{presura2000}. This temperature increase can be attributed to two possible mechanisms. First, through lattice excitation, the strongly driven IR-active phonons by THz pulse may generate acoustic phonons via phonon-phonon scattering. This scattering process gradually transfers energy from the IR-active phonons to the acoustic phonons, leading to a slow and steady increase in the lattice temperature. Among the IR-active phonons, the IR3 phonon is likely to play a particularly efficient role in this energy transfer.
A second potential mechanism involves IR-active spin excitations. Infrared spectroscopy has shown that $\alpha'$-NaV$_2$O$_5$ exhibits a magnon continuum above 139 cm$^{-1}$ (4.2 THz)~\cite{damascelli2000}. It is possible that THz light directly excites these “charged” magnons, as they are dipole-active~\cite{damascelli1999}. Consequently, through spin-phonon coupling, the energy from these excitations could then be transferred to the lattice, contributing to a rise in temperature.

The relaxation of these thermal processes leads to an overall increase in temperature, resulting in a slowly varying background in the pump-probe dynamics. This slow modulation of the reflectivity change reflects the gradual thermalization of the system following THz excitation, independent of the specific pathway, whether phonon-driven or spin-driven, that initiates the temperature increase, eventually leading to a picosecond melting of the charge order and lattice superstructure, as proposed in the photoexcitation of the low-temperature phase of $\alpha'$-NaV$_2$O$_5$~\cite{suemoto2005,nakajima2005}.

Finally, we performed a near-infrared (NIR) pump - NIR probe experiment at 5 K, merely replacing the THz stimulus with the NIR pump at a photon energy of 1.55 eV. The dynamics induced by the NIR pump are characterized by a single exponential evolution with a time constant of $\tau_{NIR} =$ 2.82 ps. Notably, this indicates that the photoinduced dynamics are slower compared to those achieved with THz excitation ($\tau_{THz}$ = 1.58 ps). This suggests that direct absorption of the THz energy through lattice vibrations or spin excitations leads to a faster increase in lattice temperature, and eventually melting charge-ordered phase, compared to the photoexcitation electronic system.
Strikingly, no significant oscillatory component is observed through optical photoexcitation. This result highlights the potential of nonlinear THz light-matter interaction in driving Raman-active phonon mode compared to purely optical pathways involving high-energy photons, such as impulsive stimulated Raman scattering~\cite{merlin1997}.\\

\section{Conclusion}
In summary, our experimental results and analytical modeling strongly suggest that the nonlinear dynamics of zone-folded Raman-active phonons in $\alpha'$-NaV$_2$O$_5$ originate from the excitation of zone-folded IR-active phonons driven to large amplitudes by intense THz pulses. This nonlinear phononic excitation process might be enhanced by the presence of charge order and superstructure formation, as suggested by Raman spectroscopy, which reveals significant lattice anharmonicity in the low-temperature phase~\cite{sherman1999}. Further studies employing ab initio density functional theory calculations could quantitatively estimate these coupling coefficients in the charge-ordered phase of $\alpha'$-NaV$_2$O$_5$ and shed light on the excitation mechanism at the microscopic level. Moreover, while disentangling electronic and lattice-driven responses remains challenging in time-resolved optical spectroscopy, ultrafast X-ray diffraction experiments at free-electron lasers offer the potential to yield crucial insights into the magnitude of coherent lattice displacements and a deeper understanding of the underlying dynamics.

In conclusion, this study extends the exploration of nonlinear THz-driven lattice dynamics to the quantum spin-ladder compound $\alpha'$-NaV$_2$O$_5$. Future investigations could focus on the dynamical modulation of the Dzyaloshinskii-Moriya (DM) interaction by THz-driven transient lattice displacements. Steady-state spectroscopic studies have already implicated lattice-mediated DM interactions in generating finite Raman and infrared cross-sections for single- and two-magnon scattering processes~\cite{room04}. By employing resonant THz-driven coherent phonons, future experiments could probe the ultrafast modulation of the infrared absorption cross-section under transient lattice displacements. Such efforts would provide novel insights into the dynamics of the DM interaction, offering experimental validation of spin-phonon coupling mechanisms that have thus far been explored primarily in steady-state regimes.

Overall, exploiting nonlinear THz light-matter interaction is emerging as an effective approach for controlling spin and lattice dynamics in strongly correlated systems. These materials  hold great potential for applications in next-generation technologies, including high-speed spintronics and THz-driven real-time data processing, paving the way for groundbreaking advances in functional device applications.\\

\textit{Acknowledgments} - The authors thank P.\ Barone for insightful discussions and gratefully acknowledge the support of the PSI-LNO GL group during the THz pump-probe experiments. FT-IR analyses were conducted at the DA$\Phi$NE-LIGHT Synchrotron Radiation Facility. The authors sincerely acknowledge the Synchrotron Radiation Service for their invaluable technical assistance with the IR spectroscopy experiments.

\bibliography{tnpbib}




\pagebreak
\onecolumngrid

\setcounter{figure}{0}
\renewcommand{\thefigure}{S\arabic{figure}}

\renewcommand{\thetable}{S\arabic{table}}

\setcounter{equation}{0}
\renewcommand{\theequation}{S\arabic{equation}}

\renewcommand{\thesubsection}{S\arabic{subsection}}
\renewcommand{\appendixname}{}

\begin{center}
\textcolor{white}{}\\
\vspace{1cm}
\large\textbf{Supplemental Material}
\end{center}
\vspace{0.cm}

\vskip8mm
\twocolumngrid
\section{Experiment setup}
\begin{figure*}[t]
\includegraphics[width=0.9\linewidth]{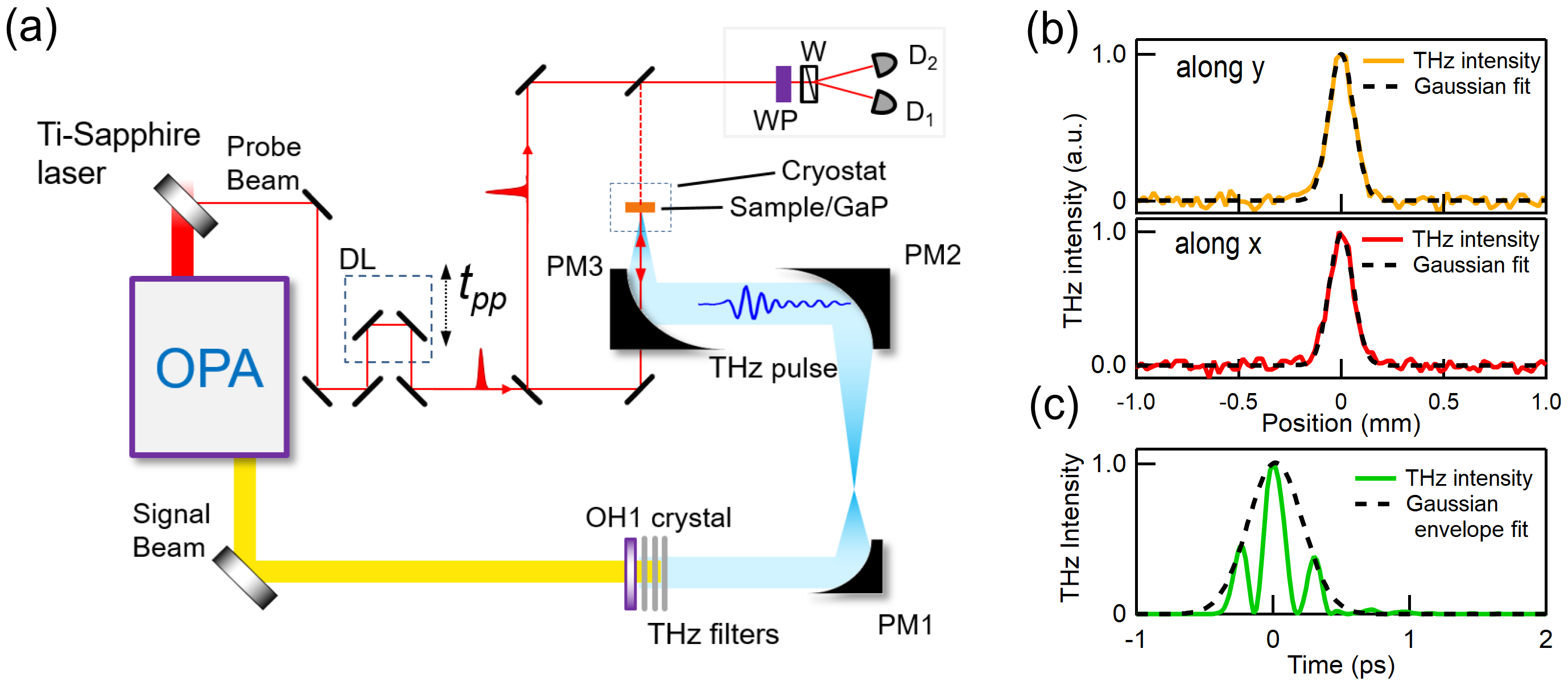}
\caption{(a) Experimental setup for THz pump NIR probe reflection spectroscopy. Optical Parametric Amplifier (OPA), parabolic mirrors (PM1-PM3), delay line (DL), wave plate (WP), Wollaston prism (W), and detectors (D1-D2). (b) THz beam profile at the sample position measured by the THz camera. Beam waists obtained by Gaussian fitting are respectively $w_x = 118$ $\mu$m and $w_y = 122$ $\mu$m (average waist $w = 120$ $\mu$m). 
(c) Temporal THz intensity waveform obtained as the square of the electric field measured by electro-optic sampling, using a Gaussian-envelope fit to determine the pulse duration.}
\label{uqmfs1}
\end{figure*}

The experimental configuration of the THz pump and optical probe setup is sketched in Fig.~\ref{uqmfs1}(a). The output of a 20 mJ, 55 fs, 800 nm Ti:Sapphire laser was used to drive an optical parametric amplifier (OPA), which provides ultrashort, multi-mJ pulses at longer wavelengths \cite{Giorgianni2019,vicario2020}. The generation of single-cycle THz pulses was based on optical rectification in organic crystal OH1 (2-[3-(4-hydroxystyryl)-5, 5-dimethylcyclohex-2-enylidene] - from Rainbow Photonics Ltd.) pumped with the OPA signal beam at 1.5 $\mu$m with an energy of 3.2 mJ.

Following the THz generation crystal, three low-pass filters were used to isolate the THz pulses and block the residual OPA beam: two of them featured a 20 THz cut-off frequency, and the third had a 6 THz cut-off frequency. These three filters gave an optical extinction ratio exceeding 10$^5$ at the OPA signal wavelength. Fig.~1(c) of the main text illustrates the resulting THz pump waveform.
Intense THz electric fields were reached by tight focusing of the THz beam using three parabolic mirrors~\cite{Giorgianni2019}.
To tailor the THz pump spectrum and obtain the four distinct pump pulses shown in Fig.~4(a)-(d) of the main text, specific spectral filters were employed. We used a 2 THz low-pass filter, a 4.2 THz low-pass filter, a 6 THz low-pass filter, and a 4.2 THz high-pass filter coupled with a 10 THz low-pass filter.

To measure NIR probe dynamics as a function of the pump strength, see Fig. 3(a) in the main text, the THz electric field was attenuated by three wire-grid polarizers. The reversal of the THz pump field polarity, depicted in Fig. 3(b) in the main text, was implemented by a 180$^\circ$ rotation of the OH1 crystal around the light propagation axis.

To estimate the electric-field strength of the pump 
pulses shown in Fig.~1 of the main text, we apply the formula~\cite{Blanchard2009}
\begin{equation}
E_{\rm THz} = \sqrt{\frac{z_0 E_p 4 \sqrt{\ln 2}}{\pi \sqrt{\pi} w^2 \tau_{\rm 
FWHM}}}, 
\label{eethz}
\end{equation}
where $z_0$ is the vacuum impedance and we measured the THz energy per pulse, $E_p = 3$ $\mu$J, using a calibrated THz energymeter (Gentec THZ12D3S-VP-D0); the beam waist, $w = 120$ $\mu$m, obtained by a Gaussian fit of the beam profile [Fig.~\ref{uqmfs1}(b)], which was measured with a micro-bolometric THz camera (NEC IRV-T0830); the pulse duration,  $\tau_{\rm FWHM} = 0.477$ ps, obtained from the FWHM of the Gaussian envelope fitting the temporal intensity waveform of the THz pump [Fig.~\ref{uqmfs1}(c)], which was taken as the square of the electric field measured at the sample position by electro-optic sampling in transmission in a 200 $\mu$m-thick (110) GaP crystal with a 50 fs, 800 nm gating pulse obtained as a fraction of the Ti:Sapphire beam. Our estimated electric-field strength, $E_{\rm THz} = 3.14$ MVcm$^{-1}$.

For the THz pump-NIR probe measurements, the gating pulse was employed to capture the ultrafast pump-induced polarization modulation of the sample in reflectivity. The THz-induced polarization modulation was measured by splitting the probe beam into two orthogonal components using a Wollaston prism.

\begin{figure}[!b]
\includegraphics[width=0.95\linewidth]{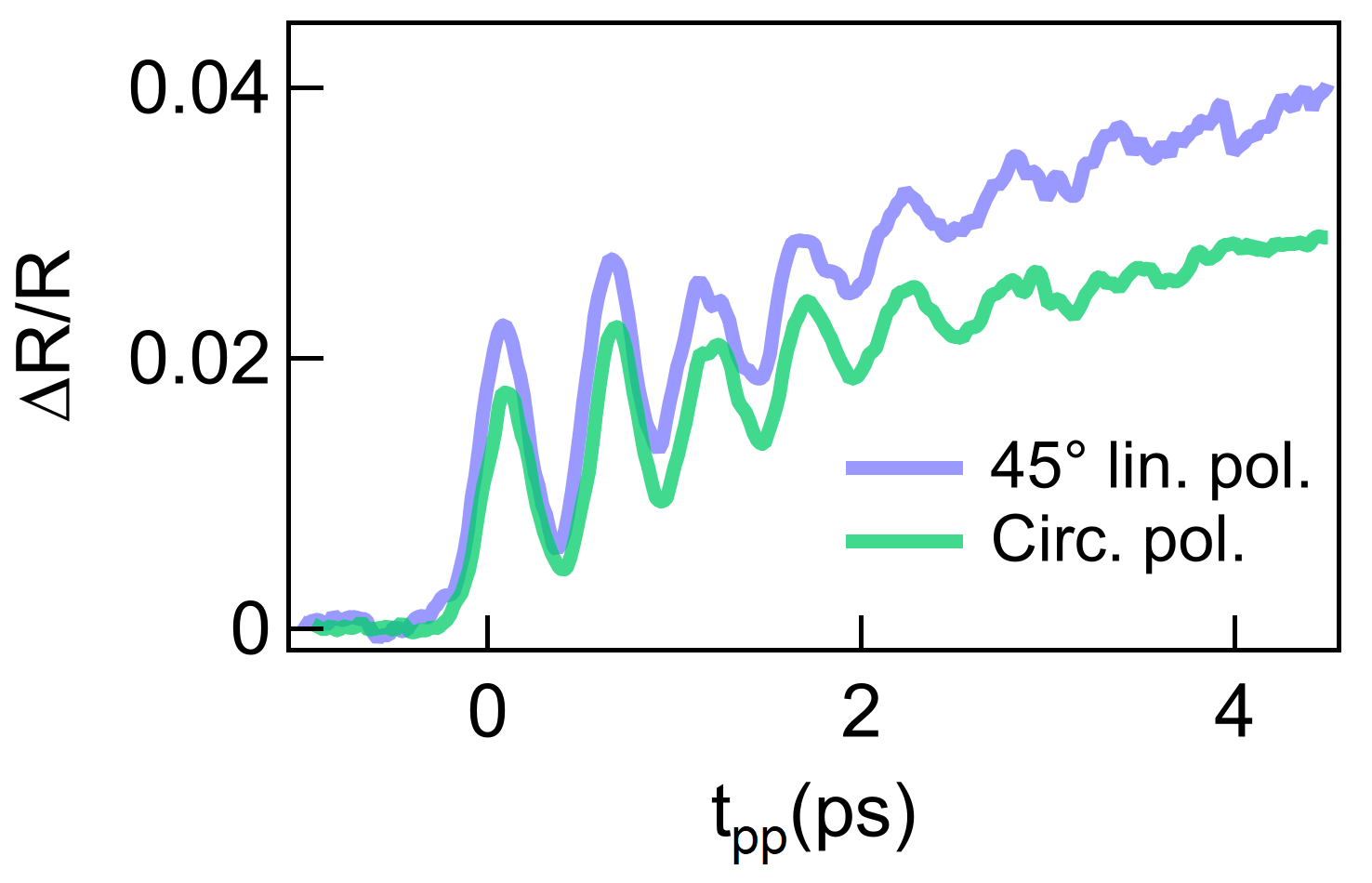}
\caption{THz pump NIR probe: comparison between circularly polarized NIR probe and 45$^\circ$ linearly-polarized NIR probe.}
\label{uqmfs2}
\end{figure}

The THz electric field of the pump was polarized in the sample plane, perpendicular to the optical table, and along the $b$ axis of the $\alpha'$-NaV$_2$O$_5$ crystal, with an uncertainty of $\pm 5^{\circ}$. The polarization of the incident probe beam sample was conventionally set to circular for a direct comparison with previous optical pump-probe experiments on $\alpha'$-NaV$_2$O$_5$. We also verify that comparable pump-probe dynamics are obtained using a 45$^\circ$-linearly polarized probe with respect to the pump polarization, see Fig.~\ref{uqmfs2}. All measurements were conducted using a He cryostat, allowing us to achieve a minimum sample temperature of 5 K.

\begin{figure*}[t]
\includegraphics[width=0.8\linewidth]{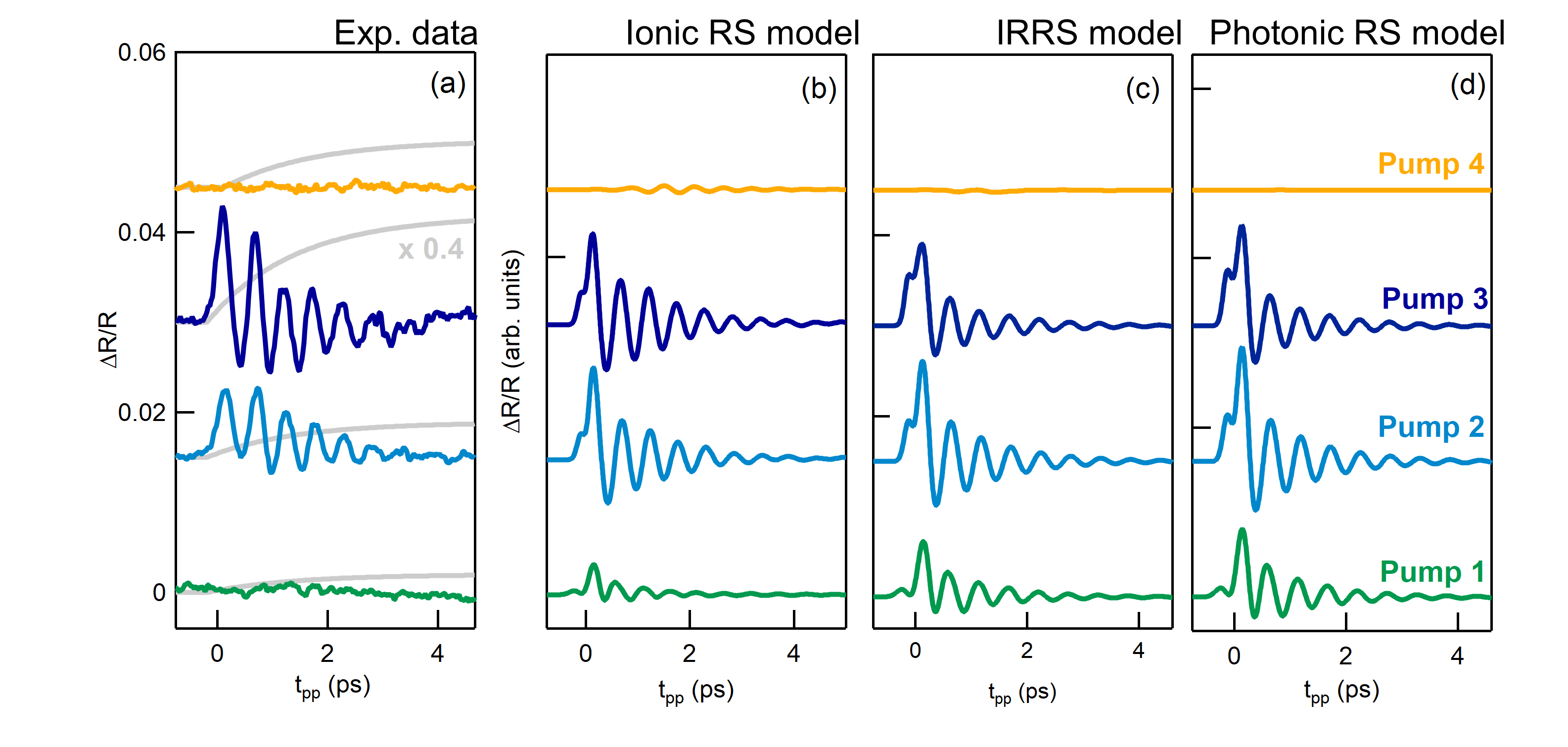}
\caption{(a) Extracted oscillatory component from THz pump-induced $\Delta R/R$ dynamics utilizing an exponential fitting procedure. The gray solid lines in these panels represent the fit using the function $\eta(t_{pp})=(1-e^{-t_{pp}/\tau})$, with a time constant $\tau_{pp}$ of 1.58 ps. Colored solid lines depict experimental data after subtracting the fit.
(b)-(d) Calculated oscillatory dynamics of $\Delta R/R$ employing the ionic RS model (b) the IRRS model (c) and the photonic RS model (d).}
\label{uqmfs3}
\end{figure*}

\section{THz-induced coeherent excitation Raman-active modes in $\alpha'$-NaV$_2$O$_5$: Nonlinear phononics \textit{vs} Nonlinear photonics}

When intense THz pulses excite the crystal lattice, distinct nonlinear pathways - phononic and photonic - can contribute to the generation of Raman-active coherent phonons~\cite{johnson2019, juraschek2018, Maehrlein2017}. In photonic pathways, the driving mechanism is governed by Raman scattering processes resonant with the sum- and difference-frequency components of the light field and commonly referred to as "photonic" Raman scattering. In phononic processes, the excitation of Raman-active modes occurs through resonant coupling to IR vibrational modes driven by intense light and includes both ionic Raman scattering and infrared resonant Raman scattering. In the first case, the coupling between phonon modes is mediated by the anharmonic lattice potential, in the latter case by the nonlinear lattice polarizability \cite{khalsa2021}.

The equation of motion for the Raman-active phonon coordinate $Q_{R}$ can be expressed as follows:

\begin{equation}\label{eqsS2}
\ddot{Q}_{R}+ 2 \gamma_{R} \dot{Q}_{R}+\Omega_R^2 Q_{R} = \mathcal{F},
\end{equation}
where $\mathcal{F}$ represents the driving force acting on the Raman-active phonon, originating from either photonic or phononic processes. Here, $\Omega_R$ is the phonon eigenfrequency, and $2\gamma_{R}$ represents the phonon linewidth. For the photonic Raman scattering (RS) mechanism, $\mathcal{F}$ is proportional to the square of the pump electric field $E(t)$: $\mathcal{F}= R E(t)^2$, being $R$ the phonon Raman tensor~\cite{juraschek2018,Maehrlein2017}. In the ionic RS process, as discussed in the main text, the driving force is given by $\mathcal{F}= \sum_{ij} c_{ijR} Q_{IR,i}Q_{IR,j}$, which arises from the ionic displacement of the IR-active phonons excited by the THz pulse, where $c_{ijR}$ represent the nonlinear coupling coefficients~\cite{juraschek2018}. The index $i,j=1, 2, 3$ indicates the IR1, IR2 and IR3 phononon modes.

In infrared resonant Raman scattering (IRRS) ~\cite{khalsa2021}, instead the driving force is $\mathcal{F}= \sum_{i} b_{i} Q_{IR,i}E(t)$, where $b_{i}$ represents the change in the effective charge of the i-th IR-active mode relative to the Raman-active phonon~\cite{khalsa2021}.

\begin{figure}[t]
\includegraphics[width=1\linewidth]{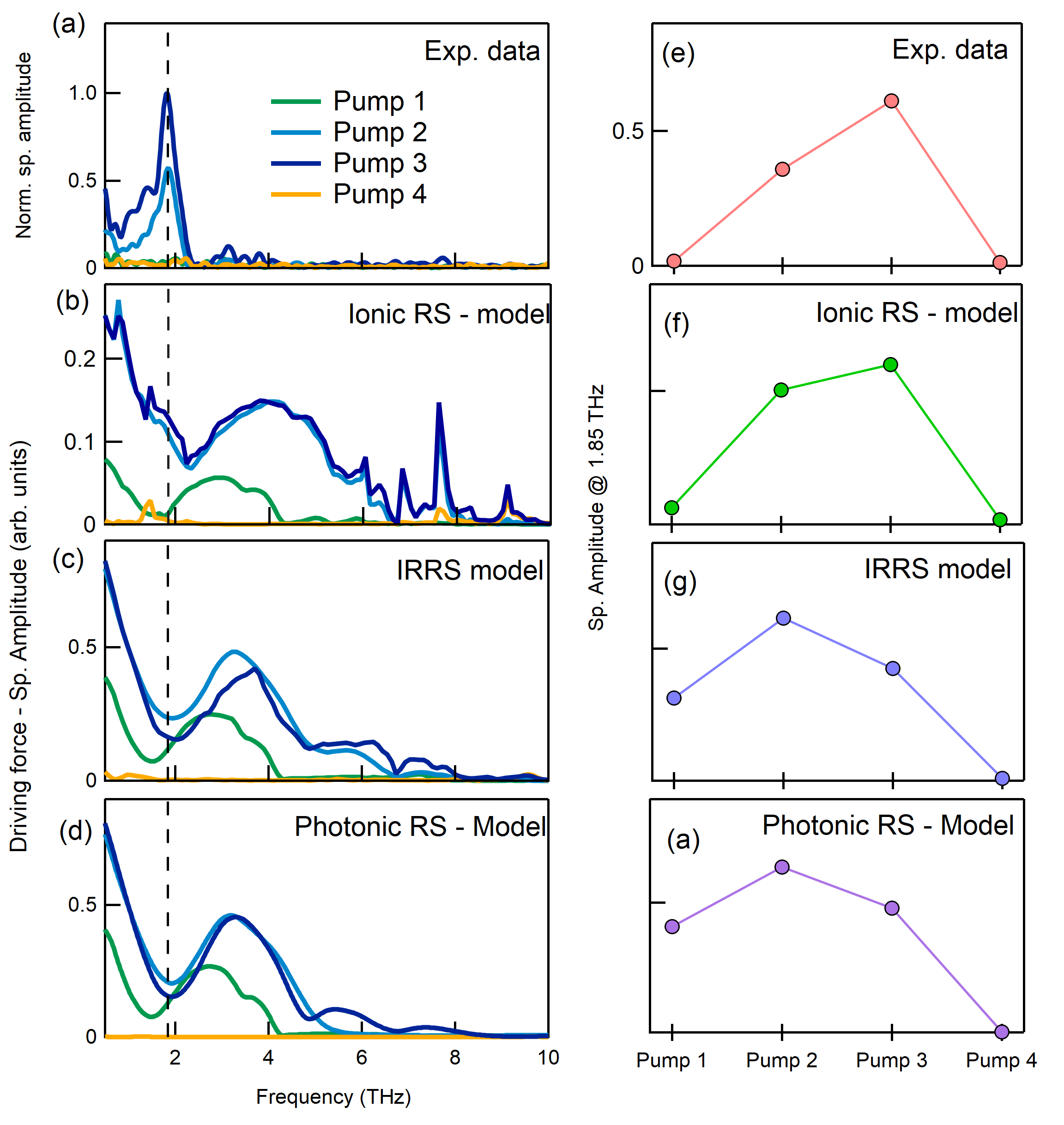}
\caption{(a) Fourier spectra of the oscillatory component of $\Delta R/R$ dynamics from Fig.~\ref{uqmfs3}(a). (b)-(d) Spectral amplitude of the calculated driving force $\mathcal{F}$ corresponding to different Raman scattering mechanisms. The black dashed lines mark the phonon resonance frequency at 1.85 THz. (e)-(h) Spectral amplitudes at $\Omega_R=$ 1.85 THz extracted from panels (a)-(d), plotted as a function of Pump 1 through Pump 4.}
\label{uqmfs4}
\end{figure}

In the following, we present a comparative analysis of the lattice dynamics modeled using ionic RS, IRRS, and photonic RS, under the THz pump excitation conditions of the experiment.

To compare these models with the experimental data, we first extract the oscillatory components from the measured pump-probe time traces (shown in Fig. 4(e)-(h) in the main text). This is achieved by isolating the oscillatory signals via subtraction of a slowly varying background by fitting the background with an exponential function $\eta(t_{pp})=c \cdot (1-e^{-t_{pp}/\tau})$, where $c$ represents the amplitude scaling factor and $\tau$ the characteristic time constant. Fig.~\ref{uqmfs3}(a) shows both the extracted oscillatory components and the exponential fits. The time constant is fixed to $\tau_{pp}$ = 1.58 ps for all the fits and $c$ is set to 0.002 for Pump 1, 0.0039 for Pump 2, 0.029 for Pump 3, and 0.029 for Pump 4.

In the numerical models, we computed the temporal dynamics of $Q_{R}$ using THz pump pulses that were experimentally measured by electro-optic sampling, see Sec. 1 for details. The spectrum of these pulses is shown in Fig. 4(a)-(c) of the main text. The oscillatory component is modeled as $\Delta R/R \propto Q_{R}$, as discussed in Ref.~\cite{huber2015}. For the ionic RS model, the phonon parameters employed in the calculation are discussed in the main text, while the coefficients $c_{ij,R}$ were phenomenologically set to have equal magnitudes due to the absence of precise information about their specific values. The results are shown in Fig.~\ref{uqmfs3}(b).

For the IRRS model, we numerically calculate the $Q_R$ dynamics with a driving force $\mathcal{F}= \sum_{i} b_{i} Q_{IR,i}E(t)$. In these calculations, we assumed equal values for the $b_{i}$ parameters due to the lack of specific information for the $\alpha'$-NaV$_2$O$_5$.

In the photonics RS model, we numerically solved Eq.~\ref{eqsS2}, with $\mathcal{F}= R E(t)^2$, as in Ref.~\cite{Maehrlein2017,juraschek2018}. The results from the photonics RS model are depicted in Fig.~\ref{uqmfs3}(d).

As shown in Fig. \ref{uqmfs3}, the amplitude of coherent oscillations in the experimentally observed dynamics is significantly suppressed when excited by a pump containing lower frequency components (Pump 1). This suppression is consistent with the ionic RS model, as the THz pump spectrum lies below the frequency of the IR-active phonons. In contrast, the $\Delta R/R$ response predicted by the IRRS and photonic RS models (Fig.~\ref{uqmfs3}) exhibits pronounced oscillatory dynamics with a significant amplitude when driven with Pump 1.

This behavior is better shown in Fig.~\ref{uqmfs4}, where the spectrum of the oscillatory component of $\Delta R/R$ experimentally measured, shown in Fig.~\ref{uqmfs4}(a), is compared with the spectrum of the computed driving force $\mathcal{F}$ for the different driving mechanisms, see Fig.~\ref{uqmfs4}(b)-(d).

The spectral amplitude at the phonon peak frequency (1.85 THz) for Pump 1 to Pump 4 is shown in Fig.~\ref{uqmfs4}(e) and compared with the spectral amplitude of the driving forces, see Fig.~\ref{uqmfs4}(f)-(h).

While the ionic RS model aligns closely with the resonant behavior observed in the experiment, the IRRS and photonic RS models do not accurately reproduce it. This reinforces the conclusion that the observed dynamics are more likely driven by nonlinear phononics, rather than the mechanisms proposed by the IRRS or photonic RS models.

\section{Model parameters}
In Table 1, we present the values of the constants $A$ and $B$ in the expression: $\Delta R/R(t_{pp})=A\cdot Q_R(t_{pp})+B\cdot\eta(t_{pp})$ for each pump in Fig. 4(i)-(n) of the main manuscript. The constant $A,$ governing the oscillation amplitude is kept unchanged for all pumps, while the constant $B$ is adjusted to reproduce the experimentally observed slowly varying background.

\begin{center}
\begin{tabular}{ |c|c|c| }
\hline
& A & B \\
\hline
Pump 1 & 70 & 3 $\cdot 10^{-3}$ \\
\hline
Pump 2 & 70 & 5 $\cdot 10^{-3}$ \\
\hline
Pump 3 & 70 & 2.8 $\cdot 10^{-2}$ \\
\hline
Pump 4 & 703& 8 $\cdot 10^{-3}$ \\
\hline
\end{tabular}
\end{center}

\section{Relation between conductivity and effective charge}

In this section, we discuss how the phonon effective charge can be estimated from the real part of the optical conductivity $\sigma_1(\omega)$. In particular, the ionic contribution to the conductivity, associated with the presence of an IR-active phonon mode, can be generally written as \cite{Bistoni2019}
\begin{equation}\label{eqs3}
\sigma^{ion}(\omega) \propto i\omega \frac{Z_{eff}^2}{(\omega+i\gamma_{IR})^2-\omega_{IR}^2},
\end{equation}
where tensor indices are neglected for simplicity. Here $Z_{eff}$ is the phonon effective charge, $\omega_{IR}$ and $\gamma_{IR}$ are the IR-active phonon frequency and effective phonon broadening, respectively. Assuming that $\gamma_{IR}\ll\omega_{IR}$, the real part of the ionic conductivity is given by 
\begin{equation}\label{eqs4}
\sigma^{ion}_1(\omega) \propto Z_{eff}^2\frac{2\gamma_{IR}\omega^2}{(\omega^2-\omega_{IR}^2)^2+4\gamma_{IR}^2\omega^2}.
\end{equation}
Since phonon peaks are usually rather sharp, $\sigma_1^{ion}$ is controlled by its $\omega=\omega_{IR}$ value, i.e.\  
\begin{equation}\label{eqs5}
\sigma_1^{ion} (\omega_{IR}) \propto \frac{Z_{eff}^2}{\gamma_{IR}}.
\end{equation}
When the electronic contribution to the conductivity can be neglected, as it happens e.g.\ in insulating compounds at frequencies below the optical gap, $\sigma_1(\omega_{IR})$ is well approximated by Eq.\ (\ref{eqs5}), and the effective charge can be estimated as:
\begin{equation}\label{eqs6}
Z_{eff}\propto \sqrt{\sigma_1 (\omega_{IR})\gamma_{IR}}.
\end{equation}

 \section{Point group and product of irreducible representations for the nonlinear phononic excitation}

In this section, we examine the symmetry properties of THz-driven phonons relevant to the nonlinear phononic excitation mechanism of the observed Raman-active mode. The exact lattice symmetry of the low-temperature charge-order phase of $\alpha'$-NaV$_2$O$_5$ remains a subject of debate. While infrared spectroscopy and Raman scattering studies have proposed that the LT phase is described either by the centrosymmetric space group \textit{C2/c} or the non-centrosymmetric Structure - \textit{Fmm2}~\cite{konstantinovic1999,popova2002}, X-ray diffraction suggests that the crystal structure may instead correspond to the monoclinic $C^3_2-A112$ space group, hosting multiple charge order configurations~\cite{XrayNVO}.
In the following, we analyze the symmetry selection rules for the nonlinear phononic excitation mechanism for all proposed space group symmetries. In the THz range of our experiment, the IR-active phonons along the $b$-axis are labeled as IR1 (at 111 cm$^{-1}$), IR2 (at 132 cm$^{-1}$), and IR3 (at 180 cm$^{-1}$). The modes IR1 and IR2 correspond to zone-folded phonons that emerge in the LT phase due to the formation of a lattice superstructure.\\

\begin{itemize}

\item \textbf{Centrosymmetric Structure - \textit{C2/c}, \textit{C2h (2/m)}:} The folded phonon modes excited by the THz field exhibit are B$_u$, with displacement coordinates confined to the $ab$ plane~\cite{konstantinovic1999,popova2002}. The observed folded Raman-active phonon has A$_g$ symmetry and, therefore, it can be detected via transient anisotropic reflectivity, as it is active in the $ab$ Raman channel~\cite{konstantinovic1999}. According to the symmetry multiplication rules, A$_g$ = B$_u$ × B$_u$. Therefore the nonlinear phononic excitation mechanism is supported by this crystal symmetry.

\item \textbf{Non-centrosymmetric Structure - \textit{Fmm2}, \textit{C2v (mm2)}:} The THz-excited IR-active folded phonon modes, IR1 and IR2, are proposed to exhibit symmetries of (2B$_1$ + 2B$_2$), making them IR-active along both the crystallographic $a$ (B$_1$) and $b$ (B$_2$) axes, as discussed in Ref.\cite{popova2002_2}. The excited IR3 phonon mode is characterized by B$_2$ symmetry and is exclusively IR-active along the $b$ axis\cite{popova2002_2,damascelli2000,room04}.
For the \textit{Fmm2} space group, the coherent phonon observed experimentally is Raman-active in the $ab$ channel and possesses A$_2$ symmetry. Notably, this mode is degenerate in frequency with a phonon mode of A$_1$ symmetry, which is also purely Raman-active but observed in the $aa$, $bb$, and $cc$ channels~\cite{popova2002}.
Since the THz pump is polarized along the $b$ axis, the IR1, IR2, and IR3 phonon modes are primarily excited with B$_2$ symmetry. Applying multiplication rules yields B$_2$ × B$_2$ = A$_1$, rather than A$_2$. Thus, multiplication rules for this crystal symmetry do not support the nonlinear phononic mechanism.

\item \textbf{Monoclinic Structure - \textit{$\mathbf{C_2^3}$-A112}:} In this case, the point group lacks the two mirror planes present in the \textit{C2v} symmetry~\cite{XrayNVO}. Consequently, the B$_1$ and B$_2$ modes, as well as the A$_1$ and A$_2$, in this lower symmetry structure transform with the same irreducible representation, $A$ and $B$ respectively. Consequently, all phonon modes in this lower-symmetry structure are predicted to exhibit both infrared (IR) and Raman activity. In the Schoenflies notation, $\mathbf{C_2^3}$ corresponds to a monoclinic cell with point group $\mathbf{C_2}$. For this point group, symmetry multiplication rules yield $B$ × $B$ = $A$, meaning that the indirect excitation of a phonon with $A$ symmetry can be obtained by driving the $B$ phonon modes. Therefore the nonlinear phononic excitation mechanism is supported by this crystal symmetry.
\end{itemize}

Beyond the ambiguities in the crystal symmetry of $\alpha'$-NaV$_2$O$_5$, charge ordering introduces further complexity. Polarization-resolved infrared spectroscopy experiments have revealed anomalies in the symmetry selection rules for zone-folded IR-active phonon modes, which have been attributed to charge ordering~\cite{room04,damascelli2000}. Specifically, it has been proposed that multiple charge-order configurations substantially alter the short-range crystal structure of $\alpha'$-NaV$_2$O$_5$, leading to local deviations from its average symmetry - a conclusion further supported by X-ray studies~\cite{xray}. This structural complexity makes the definition of selection rules in the low-temperature charge-ordered phase challenging, as well as the assignment of symmetry to the zone-folded phonons. The divergence between local and average crystal symmetries may also imply that conventional symmetry composition rules for nonlinear phononics cannot be directly applied.


\end{document}